\documentclass[
a4paper
,reprint
,amsmath
,amssymb
,superscriptaddress
,twocolumn
,aps
,pra
,nofootinbib
]
{revtex4-1}

\usepackage{graphicx}
\usepackage{dcolumn}
\usepackage{bm}
\usepackage[colorlinks]{hyperref}
\usepackage{mathtools}
\usepackage{xcolor}
\usepackage{amsthm}
\usepackage{empheq}
\usepackage{physics}
\usepackage{cleveref}

\crefname{equation}{Eq.}{Eqs.}
\crefname{figure}{Fig.}{Figs.}
\Crefname{figure}{Figure}{Figures}

\begin{document}

\title{Hierarchical Quantum Error Correction with Hypergraph Product Code and Rotated Surface Code}
\author{Junichi Haruna}
\email{j.haruna1111@gmail.com}
\affiliation{%
Center for Quantum Information and Quantum Biology, University of Osaka, 1-2 Machikaneyama, Toyonaka, Osaka 560-0043, Japan
}

\author{Keisuke Fujii}
\affiliation{%
Center for Quantum Information and Quantum Biology, University of Osaka, 1-2 Machikaneyama, Toyonaka, Osaka 560-0043, Japan
}
\affiliation{%
Graduate School of Engineering Science, University of Osaka, 
1-3 Machikaneyama, Toyonaka, Osaka 560-8531, Japan
}
\affiliation{Center for Quantum Computing, RIKEN, Hirosawa 2-1, Wako Saitama 351-0198, Japan}

\begin{abstract}
We propose and analyze a hierarchical quantum error correction (QEC) scheme that concatenates hypergraph product (HGP) codes with rotated surface codes and that is compatible with quantum computers with only nearest-neighbor interactions.
The outer code employs (3,4)-random HGP codes, known for their constant encoding rate and favorable distance scaling, while the inner code consists of a rotated surface code with distance 5, allowing hardware compatibility through lattice surgery.
To address the decoding bottleneck, we utilize a soft-decision decoding strategy that combines belief propagation with ordered statistics decoding, enhanced by a syndrome-conditioned logical error probability computed via a tailored lookup table for the inner code.
Numerical simulations under a code capacity noise model demonstrate that our hierarchical codes achieve logical error suppression below the threshold.
Furthermore, we derive explicit conditions under which the proposed codes surpass surface codes as a QEC code in both qubit efficiency and error rate. 
In particular, for the size parameter $s \geq 4$ (which corresponds to 16 logical qubits) and the distance $d\geq 25$, our construction outperforms the rotated surface code in practical regimes with physical error rates around or less than $10^{-2}$.
These results suggest that concatenated quantum low-density parity-check surface architectures can offer a scalable and resource-efficient path toward near-term fault-tolerant quantum computation.
\end{abstract}

\maketitle


\section{Introduction}
\label{sec:intro}

The surface code~\cite{Kitaev_2003,Fowler_2012} has been established as one of the most
promising quantum error correction (QEC) codes, where fault-tolerant architectures
for reliable quantum computing have been extensively investigated.
Due to its high error threshold~\cite{Dennis_2002}, robustness to local noise, and compatibility with planar architectures, the surface code has been widely adopted in both theoretical studies~\cite{Fowler_2009,Raussendorf_2007} and experimental implementations~\cite{barends2014superconducting,Takita_2017}. 
Specifically, Google recently demonstrated a quantum memory that achieves the break-even point using a variant of the surface code~\cite{acharya2024quantum}, marking a significant milestone in its practical application.

However, achieving fault-tolerant quantum computing (FTQC) with multiple logical qubits poses a substantial challenge, as it requires an enormous number of physical qubits, such as the order of $10^6$, in current architectures based on surface codes~\cite{abughanem2024ibmquantumcomputersevolution}.
A key limitation of the surface code is its low encoding rate, defined as the ratio of logical to physical qubits.
For example, the rotated surface code~\cite{Bombin_2007}, an optimized variant of the surface code, has parameters $[[L^2, 1, L]]$, where $L$ is the code distance.
Its encoding rate, $L^{-2}$, decreases as the code distance increases, resulting in inefficiencies when scaling to a large number of logical qubits.

Quantum low-density parity-check (qLDPC) codes~\cite{poulin2008iterativedecodingsparsequantum,Tillich_2014} have been proposed as an alternative to surface codes.
These stabilizer codes are characterized by two key properties that make them attractive for large-scale quantum computing:
\begin{enumerate}
    \item Each stabilizer acts on a constant number of physical qubits.
    \item Each physical qubit participates in a constant number of stabilizers.
\end{enumerate}

The high encoding rate of the qLDPC codes is expected to reduce the qubit overhead required for FTQC.
In addition, qLDPC codes exhibit favorable scaling properties, such as the code distance increasing with the number of physical qubits.
For example, hypergraph product (HGP) codes~\cite{Tillich_2014}, one of the earliest classes of qLDPC codes, achieve a constant encoding rate and a code distance scaling proportional to the square root of the number of physical qubits.
Moreover, Refs.~\cite{dinur2022goodquantumldpccodes,panteleev2022asymptoticallygoodquantumlocally} have constructed several examples of asymptotically good qLDPC codes, whose number of logical qubits and code distance both scale linearly with the number of physical qubits.
These properties make qLDPC codes a promising candidate for resource-efficient QEC.

Despite these theoretical advantages, most qLDPC codes require nonlocal qubit interactions~\cite{Bravyi_2009}, posing a significant challenge for implementation on hardware constrained to nearest-neighbor connectivity.
Although certain architectures, such as reconfigurable atom arrays~\cite{Bluvstein_2023} and multilayer layouts~\cite{delfosse2021boundsstabilizermeasurementcircuits}, can support nonlocal connections, these approaches are not directly applicable to superconducting qubit platforms.
Given the prominence of superconducting qubits in the current landscape of quantum computing, there is a pressing need to develop QEC schemes compatible with nearest-neighbor constraints.

To address this issue, we investigate a hierarchical approach in which qLDPC codes are concatenated with surface codes.
This architecture~\cite{pattison2023hierarchicalmemoriessimulatingquantum,meister2024efficientsoftoutputdecoderssurface,yoshida2024concatenatecodessavequbits,ruiz2024ldpccatcodeslowoverheadquantum,Hong_2024} utilizes the high encoding rate of qLDPC codes while inheriting the locality and hardware compatibility of the surface codes.
By being concatenated on these codes, qLDPC codes can be implemented using only nearest-neighbor interactions, making them viable for superconducting platforms.

Despite the promise of hierarchical codes, several practical challenges remain.
For example, Pattison et al.~\cite{pattison2023hierarchicalmemoriessimulatingquantum} proposed a hierarchical construction by concatenating a general constant-rate qLDPC code with the surface code, allowing 2D-local syndrome extraction.
However, their numerical estimation relied on simplified assumptions, such as hard-decision decoders or values of physical/logical error rates, leaving the question of realistic performance open.
Furthermore, decoding methods remain a major bottleneck.
Naive hard-decision decoding reduces the effective code distance of hierarchical codes.
Although soft-decision decoders such as the belief propagation decoder (BP)~\cite{Poulin_2006} can process soft-information from inner codes, they may struggle due to the degeneracy of the solution to the syndrome constraint.
Reference~\cite{meister2024efficientsoftoutputdecoderssurface} has reported that there is an error floor in the low-noise regime based only on the BP decoder, which hinders scalability.

In this work, to bridge this gap, we study conditions under which a family of the following hierarchical codes achieves better performance than surface codes as a QEC code with respect to logical error rate and qubit efficiency, using the BP decoder with a novel subroutine (ordered statistics decoding~\cite{Roffe_2020}).
As the outer code of our hierarchical construction, we adopt randomly generated HGP codes with $25s^2$ physical qubits and $s^2$ logical qubits, where $s$ is an integer size parameter.
These codes are concatenated with a rotated surface code of size $L = 5$ of the inner code.
Then, we evaluate the logical error rates under a code capacity noise model and compare the performance of our hierarchical codes to that of the rotated surface code.

For decoding, we employ soft-decision decoding for the HGP code using the belief propagation--ordered statistics (BP-OS) decoder~\cite{Roffe_2020}.
For the inner code, we construct a lookup-table decoder for the $L=5$ rotated surface code.
A novel advantage of constructing this lookup table is that it allows us to compute syndrome-conditioned logical error probabilities {\it rigorously}, which are then used in the decoding of the outer code.

Finally, we explore the parameter regime in which our hierarchical code outperforms the rotated surface code as a QEC code in terms of logical error suppression and qubit efficiency.
Our findings show that when the physical error rate is around or less than $10^{-2}$, our hierarchical code achieves superior performance with $k \geq 16$ logical qubits and the code distance $d \geq 25$.
These results suggest that hierarchical QEC provides a potential pathway toward scalable and resource-efficient FTQC.

This paper is organized as follows. In \Cref{sec:preliminary}, we review the construction and key properties of the HGP codes and code concatenation.
\Cref{sec:simulation_setup_and_performance_evaluation} details the numerical setup and results, including the analysis of the logical error rate of our concatenated codes with soft-decision decoding.
Then, we discuss the conditions under which our concatenated codes outperform the rotated surface code as a QEC code in terms of qubit efficiency and logical error rates.
Finally, \Cref{sec:summary_outlook} summarizes our findings and describes future directions.

\section{Preliminary}
\label{sec:preliminary}

In this paper, we study the hypergraph product (HGP) code concatenated with the rotated surface code. 
This section provides a brief review of the HGP code and the concept of code concatenation.

\subsection{Hypergraph Product Code}
\label{subsec:hgp}

The hypergraph product (HGP) code~\cite{Tillich_2014} is one of the earliest examples of qLDPC codes.
It is constructed from two classical linear codes and inherits the LDPC property when both classical codes satisfy the LDPC conditions.

A notable advantage of the HGP code is its simple construction as a Calderbank--Shor--Steane (CSS) code~\cite{Calderbank_1996,shor1997faulttolerantquantumcomputation}, where stabilizers are formed using the Pauli-$X$ and Pauli-$Z$ operators.
Given two classical linear codes with parity-check matrices $H_1$ and $H_2$ of sizes $m_1 \times n_1$ and $m_2 \times n_2$, respectively, the HGP code contains $n_1 n_2 + m_1 m_2$ physical qubits.
The stabilizer generators, denoted by $H_X$ and $H_Z$, are given in the binary representation by:
\begin{subequations}
\label{eq:definition_of_parity_check_matrices_hgp}
\begin{align}
    H_X &= \begin{bmatrix} H_1 \otimes I_{n_2} & I_{m_1} \otimes H_2^T \end{bmatrix}, \\
    H_Z &= \begin{bmatrix} I_{n_1} \otimes H_2 & H_1^T \otimes I_{m_2} \end{bmatrix}.
\end{align}
\end{subequations}
The parameters of the HGP code are expressed as:
\begin{align}
    [[n_1 n_2 + m_1 m_2, k_1 k_2 + k_1^T k_2^T, \min(d_1, d_2, d_1^T, d_2^T)]],
\end{align}
where
\begin{itemize}
    \item $k_i$: number of logical bits in the classical code corresponding to $H_i$, defined as $\dim(\ker H_i)$.
    \item $k_i^T$: number of logical bits in the transposed matrix $H_i^T$.
    \item $d_i$: code distance of the classical code, defined as:
    \begin{align}
        d_i = \min_{\psi \in \ker H_i\cap \psi\neq \vec{0}} \mathrm{wt}(\psi),
    \end{align}
    where $\mathrm{wt}(\psi)$ is the Hamming weight of $\psi$. If $k_i = 0$, $d_i$ is set formally to $\infty$.
    \item $d_i^T$: code distance for the transposed matrix $H_i^T$.
\end{itemize}

\subsubsection*{Advantages of HGP Codes}
The HGP codes offer several advantages as QEC codes.
Firstly, the LDPC property is preserved because the stabilizer weight is determined by the row and column weights of the classical parity-check matrices.
For simplicity, consider the case where $H_1 = H_2$ with $a \coloneqq \max_i \sum_j (H_1)_{ij}$ and $b \coloneqq \max_j \sum_i (H_1)_{ij}$.
The stabilizer weight of the HGP code is then bounded by $a + b$, which remains constant if $a$ and $b$ are independent of the size of the classical parity-check matrix.

Another significant advantage is the high encoding rate.
Suppose that the classical code has a constant encoding rate, i.e. $k_1 = r \cdot n_1$ for some $r \, (0 < r < 1)$.
In this case, $H_1$ becomes a matrix of size $n_1 \times (1-r)n_1$, and the number of logical bits for $H_1^T$ is $k_1^T = 0$.
Consequently, the HGP code has parameters:
\begin{align}
    [[(1 + (1-r)^2)n_1^2, r^2 n_1^2, d_1]].
\end{align}
The encoding rate, $r^2 / (1 + (1-r)^2)$, remains constant as the number of physical qubits increases, contrasting with the diminishing encoding rate of surface codes.

If the classical code has the code distance $d_1$ proportional to the number of bits $n_1$, then that of the resulting HGP code behaves as the square root of the number of physical qubits ($d_1 \propto \sqrt{n_1^2}$), as discussed in the original paper~\cite{Tillich_2014}.
Furthermore, if we construct HGP codes with randomly generated classical parity-check matrices, the scaling of their code distance depends on the classical codes.
This point will be discussed later in \Cref{subsec:random_hgp}.

\subsubsection*{Challenges of HGP Codes}
Despite these benefits, the HGP codes face challenges in their practical implementation.
Specifically, the stabilizers often involve nonlocal interactions between qubits, which are difficult to realize on planar architectures such as superconducting qubits.
To address this, various methods have been proposed, including multilayer architectures~\cite{delfosse2021boundsstabilizermeasurementcircuits} and reconfigurable atom arrays~\cite{Bluvstein_2023}.

In this paper, we address these challenges employing a code concatenation approach, which is reviewed in the next subsection.
This strategy enables the implementation of the HGP codes using only nearest-neighbor interactions, making them suitable for practical quantum hardware.

\subsection{Code Concatenation}
\label{subsec:concatenation}

Code concatenation~\cite{knill1996concatenatedquantumcodes} provides a practical solution to the nonlocality issue by enabling the implementation of qLDPC codes using only nearest-neighbor interactions.
In this approach, logical qubits are encoded across multiple layers of QEC codes.

Consider a concatenated code formed by encoding a quantum code $\mathcal{Q}_2$ with parameters $[[n_2, k_2, d_2]]$ on top of another quantum code $\mathcal{Q}_1$ with parameters $[[n_1, k_1, d_1]]$.
The concatenation process involves:
\begin{itemize}
    \item Encoding $k_1$ logical qubits into $n_1$ physical qubits by the inner code $\mathcal{Q}_1$.
    \item Treating the logical subspace of each $\mathcal{Q}_1$ block as physical qubits for the outer code $\mathcal{Q}_2$.
    \item Encoding $k_2$ logical qubits using $\mathcal{Q}_2$ across the $n_2$ physical qubits encoded by $\mathcal{Q}_1$.
\end{itemize}
The resulting concatenated code has parameters:
\begin{align}
    [[n_1 n_2, k_1 k_2, d_1 d_2]].
\end{align}

\subsubsection*{Advantages of Code Concatenation}
One key advantage of code concatenation is the exponential suppression of logical error rates with the number of redundant physical qubits.
For example, if a quantum code with parameters $[[n, 1, d]]$ is concatenated $N$ times, the logical error rate $p_L$ behaves in the low-error regime as:
\begin{align}
    p_L \sim \qty(\frac{p}{p_\mathrm{th}})^{(\lfloor \frac{d+1}{2} \rfloor)^ N},
\end{align}
where $p_\mathrm{th}$ is the threshold error rate, independent of $N$.
When the physical error rate $p$ is below this threshold, arbitrarily low logical error rates can be achieved by increasing the number of concatenation levels sufficiently.
This forms the basis of the threshold theorem~\cite{aharonov1999faulttolerantquantumcomputationconstant,Knill_1998,Kitaev_2003,shor1997faulttolerantquantumcomputation}.

\subsubsection*{Decoding Strategies}
Decoding in concatenated codes involves estimating errors based on syndrome measurements.
Two commonly used strategies are hard-decision decoding and soft-decision decoding~\cite{Poulin_2006}.

{\it Hard-Decision Decoding:}
This approach simplifies the decoding process but cannot fully utilize the code distance of concatenated codes.
It involves:
\begin{enumerate}
    \item Decoding each inner code block using syndrome extraction, and applying recovery operations at the inner code.
    \item Performing syndrome extraction for the outer code, followed by decoding of the outer code based on these syndromes.
\end{enumerate}
Although computationally efficient, this method reduces the effective code distance to:
\begin{align}
    d_{\mathrm{eff}} = 2\left\lfloor \frac{d_1+1}{2} \right\rfloor \left\lfloor \frac{d_2+1}{2} \right\rfloor,
\end{align}
as logical errors in inner codes propagate to the outer code as physical errors.

{\it Soft-Decision Decoding:}
This approach uses additional information, such as syndrome-dependent probabilities from decoding of the inner code, to improve error estimation of the outer code.
Soft-decision decoding achieves better scaling of the logical error rate and could utilize the full code distance $d_1 d_2$~\cite{Poulin_2006}.
While computationally more demanding, it provides significant performance improvements, particularly for large-scale concatenated codes.

\begin{figure}[t]
    \centering
    \includegraphics[width=0.3\textwidth]{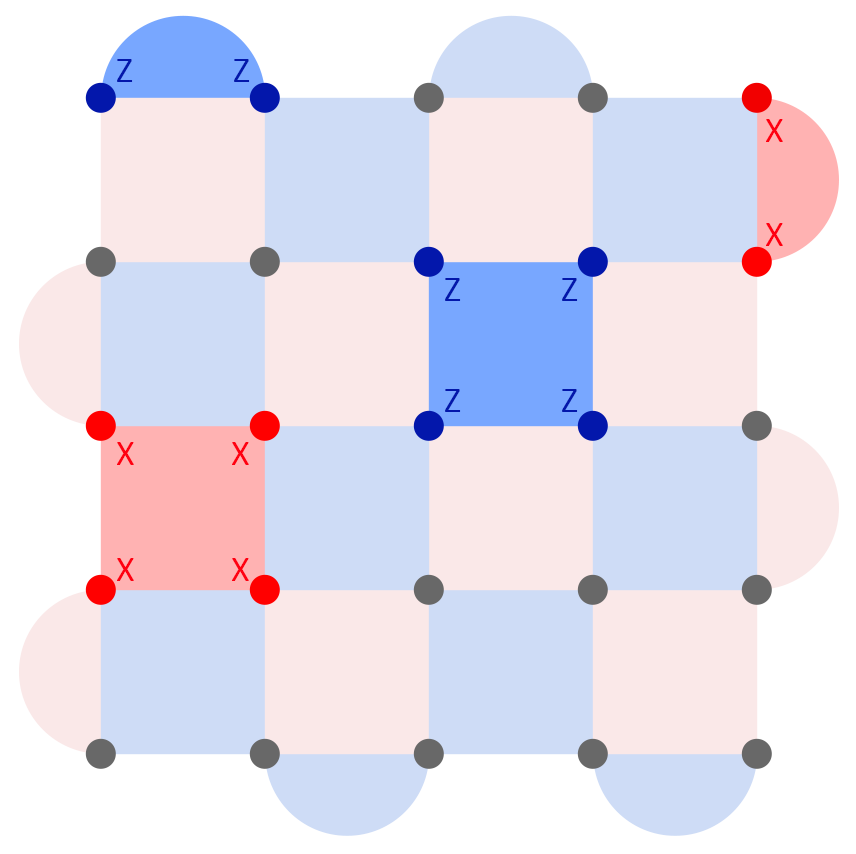}
    \caption{Illustration of the rotated surface code of the size $L=5$.
    The red or blue areas correspond to X or Z stabilizers, respectively.}
    \label{fig:Illustration_of_rotated_surface_code}
\end{figure}

\subsubsection*{Surface Code as the inner code and Lattice Surgery}

The surface code~\cite{Kitaev_2003,Fowler_2012} is a kind of topological QEC code, widely recognized for its fault-tolerance properties and compatibility with 2D quantum architectures. 
An illustration of the rotated surface code is shown in \cref{fig:Illustration_of_rotated_surface_code}. 
In this code, physical qubits are arranged on a 2D square lattice, where stabilizer measurements are performed using four- or two-body Pauli-$X$ and Pauli-$Z$ operators acting on sets of nearest-neighbor qubits. 
A single logical qubit is encoded regardless of the code size. 
Logical $X$ and $Z$ operations correspond to the product of Pauli operators forming a connected path across the lattice linking opposite boundaries.

When the surface code is used as the inner code in concatenated codes, the lattice surgery technique~\cite{Horsman_2012,Fowler_2012,Litinski_2019} can be used to perform logical operations.
The lattice surgery involves manipulating the boundaries of surface code patches, enabling operations such as merging or splitting patches to perform logical gates.
This boundary-based approach eliminates the need for direct physical interactions between distant qubits.

The primary advantage of the lattice surgery in concatenated code architectures lies in its compatibility with planar hardware, such as superconducting qubits arranged in a 2D grid.
By utilizing it, nonlocal connectivity requirements for outer codes (e.g. the HGP codes) can be effectively realized, reducing qubit overhead and simplifying the implementation of logical gates across surface code blocks.

\section{Performance Evaluation of Hierarchical Hypergraph Product Code}
\label{sec:simulation_setup_and_performance_evaluation}

\begin{figure*}[ht]
    \centering
    \includegraphics[width=0.7\textwidth]{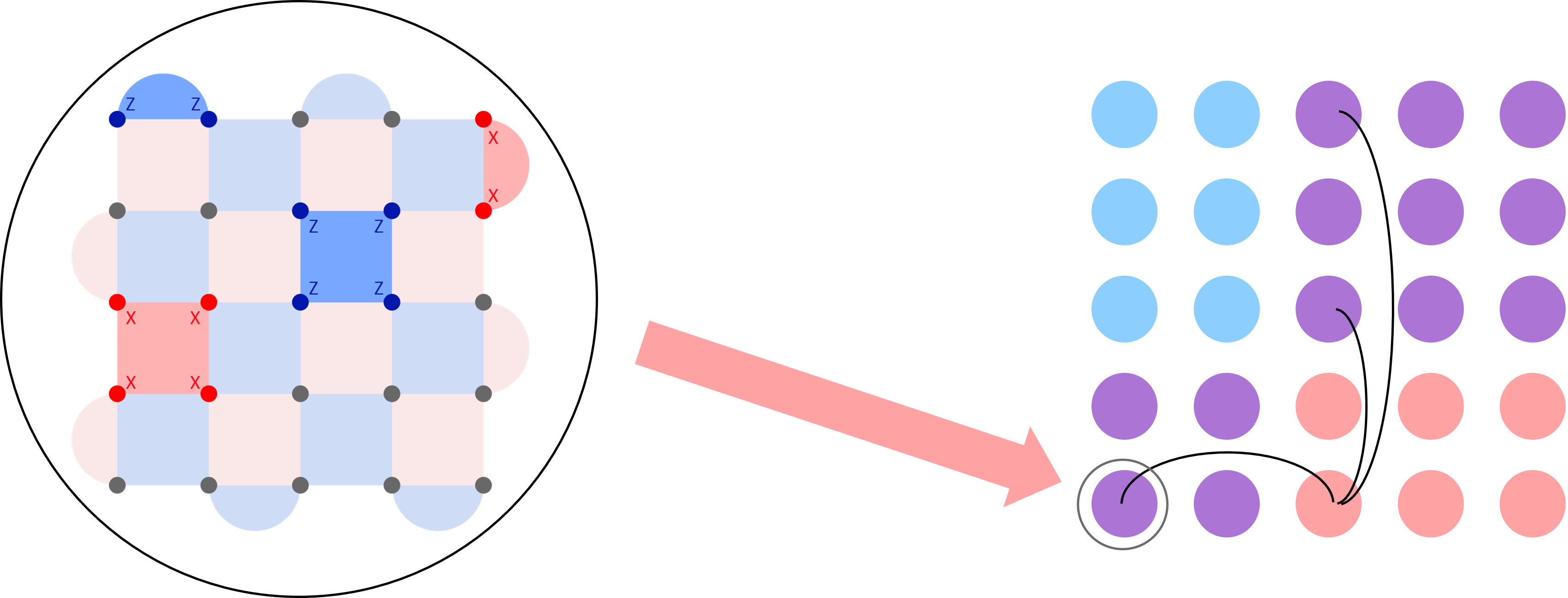}
    \caption{Illustration of the concatenated code: the inner code uses a rotated surface code, and the outer code employs an HGP code.}
    \label{fig:Illustration_of_the_concatenated_code}
\end{figure*}

In this section, we explain the setup of our concatenated code and analyze its performance as a QEC code compared to the rotated surface codes.

The architecture of the concatenated code is illustrated in \cref{fig:Illustration_of_the_concatenated_code}. 
The inner code consists of a rotated surface code of size $L=5$, chosen for its compatibility with lattice surgery and planar hardware layouts. 
The outer code employs HGP codes constructed from randomly generated classical linear codes of various sizes. 
The resulting concatenated code is described by the parameters:
\begin{align}
    [[625s^2, s^2, 5d_s]],
\end{align}
where $s$ is the size parameter of the classical parity-check matrix, and $d_s$ represents the largest code distance observed among numerically sampled instances for a given $s$.
These parameters are analyzed in detail in later subsections.

Here we briefly explain the reason for selecting the rotated surface code with size $L = 5$ as the inner code and the (3,4)-random HGP codes as the outer code.
The inner code was chosen primarily for its compatibility with lattice surgery.
Additionally, we used a lookup-table decoder for the inner code to enable rigorous calculation of syndrome-conditioned logical error probabilities.
Since the number of possible $X$ or $Z$ error configurations scales exponentially as $2^{L^2}$, practical memory limits constrain the code size that can be handled in this way.
For instance, using $L=7$ would require over 70 terabytes of memory to store the full lookup table.
As a compromise between decoding feasibility and error suppression capability, we considered $L=3$ and $L=5$ and ultimately selected the latter for its better performance.
A more detailed discussion of the inner code size is provided in \Cref{app:generalization_to_any_rotated_suface_code}.

For the outer code, we adopt the (3,4)-random HGP code.
While the general construction and properties of HGP codes are reviewed in \Cref{sec:preliminary}, we emphasize here that their simple and nice geometrical structure is advantageous for hardware implementation and decoder design.
Owing to this feature, HGP codes have been extensively studied in the context of syndrome extraction, stabilizer measurement circuits, and logical gate implementations~\cite{Roffe_2020,delfosse2021boundsstabilizermeasurementcircuits,quintavalle2022reshapedecoderhypergraphproduct,Quintavalle_2023,xu2024fastparallelizablelogicalcomputation}.
Among various $(a,b)$-random variants, the (3,4)-random HGP code has a favorable balance between stabilizer weight and encoding rate.
In particular, it achieves the highest known encoding rate of $1/25$ while preserving code distance scaling proportional to $\sqrt{n}$, where $n$ is the number of physical qubits, as will be discussed in the following subsection.
These features make the (3,4)-random HGP code a practical and scalable choice for the outer code in our hierarchical construction.

The purpose of this study is to assess the logical error rates of the concatenated codes and determine the minimum size parameter $s$ required for the concatenated code to outperform the rotated surface code in terms of error suppression and qubit efficiency. 
Specifically, in \Cref{subsec:random_hgp}, we analyze the code distance and number of logical bits of the classical codes used to construct the HGP codes.
Next, in \Cref{subsec:logical_error}, we evaluate the logical error rates of the concatenated codes under a code capacity noise model using Monte Carlo simulations.
Additionally, we perform a fitting analysis to extract the error scaling factor and pseudo-threshold for each size parameter.
The size dependence of the scaling factor is also examined numerically.
Finally, in \Cref{sec:How_Large_Hierarchical_HGP_Code_Can_Outperform_Rotated_Surface_Code}, we discuss how large concatenated code can outperform the rotated surface code in terms of qubit efficiency and logical error rate, based on numerical results.

\subsection{Randomly Generated HGP Codes}
\label{subsec:random_hgp}

To construct the HGP codes, we generate classical linear codes following the approach in Ref.~\cite{delfosse2021boundsstabilizermeasurementcircuits}. 
The classical parity-check matrix $H_c$ is designed to satisfy fixed row and column weights:
\begin{align}
    \sum_i (H_c)_{ij} = 3, \quad \sum_j (H_c)_{ij} = 4,
\end{align}
ensuring that each column has weight 3 and each row has weight 4. 
The dimensions of the matrix are $3s \times 4s$, where $s$ is the size parameter. 
The HGP codes derived from these matrices are called (3,4)-random HGP codes.
We refer to HGP codes constructed with the classical parity-check matrix where each column has weight $a$ and each row has weight $b$ as $(a,b)$-random HGP codes, in general.

The choice of (3,4)-random codes is guided by the trade-off between the encoding rate and the weight of the stabilizer. 
For an $(a,b)$-random code, the classical parity-check matrix $H_c$ is a binary matrix of size $a\cdot s \times b\cdot s$, with the number of logical bits given by $k_c = (b-a)s$ and $k_c^T = 0$ generically. 
The corresponding HGP code, constructed solely from this classical code, has parameters:
\begin{align}
    [[(a^2+b^2)s^2, (b-a)^2s^2, d_c]].
\end{align}
Maximizing the encoding rate requires optimizing the ratio $(b-a)^2 / (a^2 + b^2)$, which favors larger values of the ratio $b/a$. 
However, increasing the stabilizer weight $(a+b)$ typically leads to vulnerability to fault tolerance, making the measurement of the syndrome complicated.
Then, we have to consider the balance between encoding efficiency and error correction performance.
Among the various configurations evaluated, (3,4)-random codes achieved a good balance between these factors.
In contrast, $(2,m)$-random codes such as (2,3) and (2,4) have a logarithmic scaling code distance ($d \sim \log s$) as $s$ increased~\cite{gallager_low-density_1963}, making them unsuitable for scalable QEC.

\begin{figure}[htbp]
    \centering
    \includegraphics[width=1.0\linewidth]{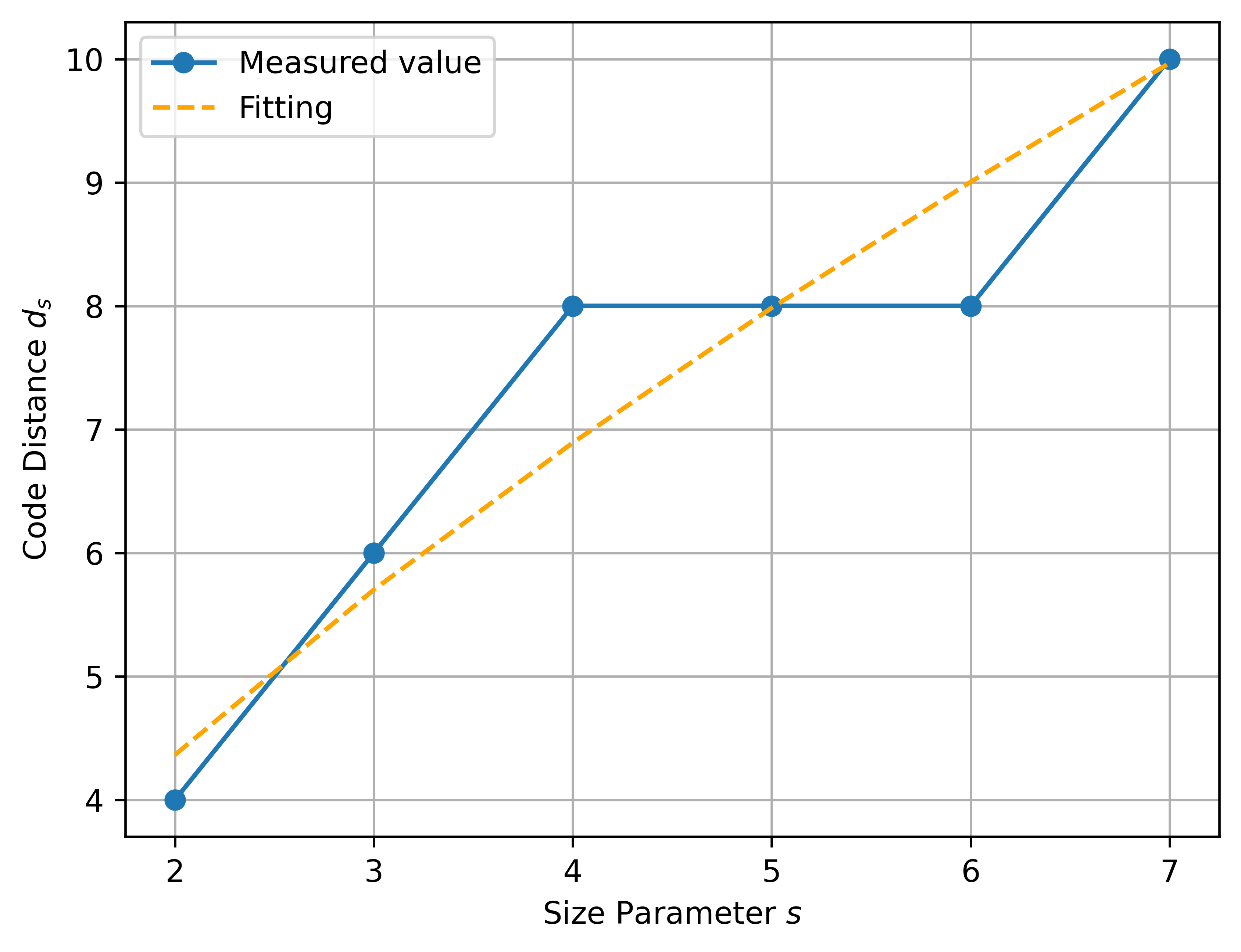}
    \caption{Maximum observed code distance $d_s$ of (3,4)-random HGP codes as a function of the size parameter $s$ (blue line).
    The orange line depicts the fitting line given by \Cref{eq:scaling_of_code_distance}.
    Data points represent the highest code distance found among 1000 randomly generated samples for each $s$. The observed growth of $d_s$ with increasing $s$ demonstrates the scalability of HGP codes.}
    \label{fig:relation_between_size_and_distance}
\end{figure}

For each size parameter $s$, we generated 1000 random instances of $H_c$ and identified the largest observed code distance $d_s$ among them. 
The results are summarized in \cref{fig:relation_between_size_and_distance}.
For later use, we perform a fitting analysis of the code distances against the size parameter.
Assuming the following formula $d_s = b_h \cdot s^{c_h}$, we get
\begin{align}
\label{eq:scaling_of_code_distance}
b_h = 2.76, c_h = 0.660.
\end{align}
This behavior (\cref{fig:relation_between_size_and_distance}) is consistent with the code distances for (3,4)-random codes in the previous literature~\cite{Roffe_2020}.

We also analyzed the number of logical bits, $k_c$ and $k_c^T$, for the parity-check matrices with the largest code distance observed at each $s$. 
Our results indicate that these values are given by:
\begin{align}
    k_c = s, \quad k_c^T = 0.
\end{align}
For subsequent evaluations, we selected the parity-check matrices corresponding to $s=2,3,4,5,6$ to construct HGP codes. 
The resulting HGP codes have parameters:
\begin{align}
    [[25s^2, s^2, d_s]].
\end{align}
These HGP codes serve as the outer codes in our concatenated code construction.

\subsection{Logical Error Rate of Concatenated Codes}
\label{subsec:logical_error}

To evaluate the performance of the concatenated code, we performed Monte Carlo simulations under the code capacity noise model; each data qubit is subject to a depolarizing noise model, as outlined below.

\subsubsection*{Setup and Decoding Strategy}
\label{subsubsec:monte_carlo}

The noise channel of the depolarizing noise model is defined as:
\begin{align}
    \mathcal{E}_p(\rho) = (1-p)\rho + \frac{p}{3}(X\rho X + Y\rho Y + Z\rho Z),
\end{align}
where $\rho$ is a density matrix and $p$ represents the physical error rate. 
In our analysis, we do not distinguish between different types of logical Pauli errors ($X, Y, Z$); instead, we consider the total logical error rate as the sum of all logical Pauli error rates. 

Since the concatenated code encodes multiple logical qubits, each logical qubit may exhibit slightly different error scaling behavior.
However, we have confirmed that these differences are minor and that the logical error rates follow similar trends across different logical qubits.
Therefore, we here report the average logical error rate over all logical qubits in the concatenated code.

The decoding procedure consists of two stages:
\begin{itemize}
    \item \textbf{Inner code:} The rotated surface code is decoded using a lookup-table decoder.
    \item \textbf{Outer code:} The HGP code is decoded using the belief propagation--ordered statistics (BP-OS) decoder~\cite{Roffe_2020} of depth of $\lambda = 10$, which supports soft-decision decoding~\cite{Poulin_2006}.
\end{itemize}

In this study, we used soft-decision decoding due to its superior error suppression capabilities. 
Unlike hard-decision decoding, which collects failure probabilities across all syndromes, soft-decision decoding requires syndrome-dependent failure probabilities to improve decoding accuracy. 
A key challenge in hierarchical quantum error correction is the calculation of these syndrome-conditioned logical error probabilities. 
Most decoders only output an expected error configuration for a given measured syndrome but do not provide the probability distribution over different error configurations. 
Although several soft-output decoders for the surface code have been proposed~\cite{meister2024efficientsoftoutputdecoderssurface}, we construct a lookup-table decoder tailored to our setup.

Here, we assume a depolarizing noise model and exploit the fact that the rotated surface code is a CSS code.
Under this assumption, the syndromes corresponding to $X$ and $Z$ errors can be decoded independently.
For a rotated surface code with $L=5$, there are $2^{25} \approx 3.4 \times 10^7$ possible error configurations for $X$ or $Z$ errors. 
Since classifying all these error configurations based on their syndromes is computationally feasible, we compute the lookup table before running Monte Carlo simulations.
More specifically, we can construct a set of error configurations $V(s)$ that gives a syndrome $s$ as
\begin{align}
    V(s) \coloneqq \{ e \in \mathbb{F}_2^n | He = s\},
\end{align}
where $H$ is the binary representation of the $X$ or $Z$ stabilizers for $Z$ or $X$ errors $e$, respectively.
$s$ is the binary vector of the syndrome and $\mathbb{F}_2$ is the finite field with two elements.

For each syndrome, we determine the most probable error configuration $e_p(s)$ by selecting the one with the minimum Hamming weight:
\begin{align}
    e_p(s) \coloneqq \operatornamewithlimits{argmin}\limits_{e \in V(s)}(\mathrm{wt}(e)).
\end{align}
This $e_p(s)$ gives the recovery operation for a given syndrome $s$ if the physical error probability is less than half $p < 1/2$.
These $V(s)$ and $e_p(s)$ allow us to specify the error configurations $V_e(s)$ that lead to a logical error after recovery operation:
\begin{multline}
    V_e(s) \coloneqq
    \\ \{ e \in V(s) | {}^\exists L \text{ s.t. } L \cdot (e + e_p(s)) = 1\, (\text{mod } 2)\},
\end{multline}
where $L$ is the binary representation of (one of) the logical $X$ or $Z$ operator for $Z$ or $X$ error $e$, respectively.
Simultaneously, we can find a set $V_c(s)$ of correctable errors for a given syndrome $s$:
\begin{align}
    V_c(s) \coloneqq V(s) - V_e(s).
\end{align}
By leveraging this precomputed information $V_e(s)$ and $V_c(s)$, we can {\it rigorously} compute the syndrome-conditioned logical error probability $P_L(p|s)$ by
\begin{align}
\label{eq:definition_of_syndrome-conditioned_logical_error_probability}
    P_L(p|s) \coloneqq \frac{P_e(p,s)}{P_e(p,s) + P_c(p,s)},
\end{align}
where 
\begin{align}
    P_e(p,s) &\coloneqq \sum_{e\in V_e(s)} p^{\mathrm{wt(e)}}(1-p)^{\mathrm{wt(e)}}, \\
    P_c(p,s) &\coloneqq \sum_{e\in V_c(s)} p^{\mathrm{wt(e)}}(1-p)^{\mathrm{wt(e)}},
\end{align}
and $p$ is the physical error probability.
This $P_L(p|s)$ is then used as an initial probability of the BP decoder to improve the accuracy of soft-decision decoding in the outer HGP code.

As for decoding of the outer HGP code, we utilize the belief propagation--ordered statistics (BP-OS) decoder~\cite{Roffe_2020,Panteleev_2021}.
This decoder operates in two stages.
First, the belief propagation (BP) algorithm estimates the marginal error probabilities for each qubit through iterative message passing.
If the resulting estimate fails to reproduce the observed syndrome, the decoder invokes the ordered statistics (OS) decoding procedure as a post-processing step.
The OS algorithm identifies a set of highly reliable bits based on the BP marginals, constructs a full-rank submatrix of the parity-check matrix using those bits, and computes a candidate solution by inverting this submatrix.
To further improve decoding accuracy, a small number of additional bit-flip configurations are explored within a restricted subspace, where the number of flips is limited by a predefined depth parameter (called the OSD order).
In this work, we apply BP-OS decoding with depth $\lambda = 10$.

The BP-OS decoder has been shown to perform well across a wide range of qLDPC codes constructed from hypergraph products.
Notably, its threshold performance for the toric code is comparable to that of the minimum-weight perfect matching (MWPM) decoder, and it achieves exponential suppression of logical errors in the low-error regime for families of random or semi-topological qLDPC codes.
For these reasons, BP-OS decoding is widely regarded as a standard and effective decoder for qLDPC codes.
Importantly, the BP stage requires prior knowledge of the marginal error probabilities for each qubit.
In our hierarchical decoding scheme, we supply this information using the syndrome-conditioned logical error probability $P_L(p|s)$ defined in \Cref{eq:definition_of_syndrome-conditioned_logical_error_probability}, obtained from the lookup-table decoder for the inner code.
This enables us to implement a soft-decision decoding strategy for the outer HGP code, leading to improved decoding accuracy.

\subsubsection*{Logical Error Rate Results and Fitting Analysis}
\label{subsubsec:logical_error_results}

\begin{figure}[ht]
    \centering
    \includegraphics[width=1.0\linewidth]{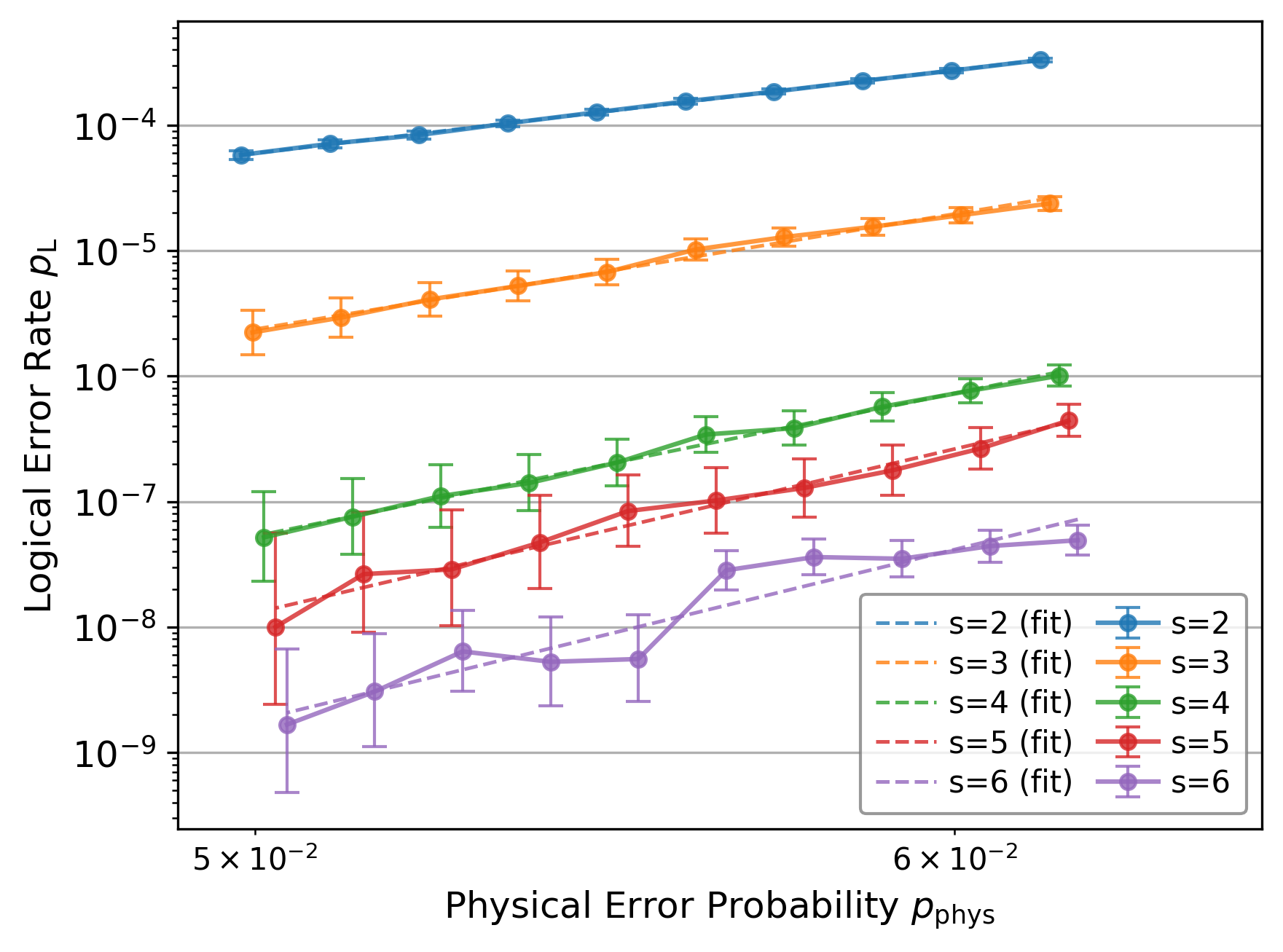}
    \caption{
            Logical error rates of concatenated HGP codes as a function of the physical error probability $p_{\mathrm{phys}}$, for size parameters $s = 2, 3, 4, 5, 6$. 
            Logical error rates are defined as the sum of logical $X$, $Y$, and $Z$ error rates, averaged over multiple logical qubits. 
            Monte Carlo simulations were performed with $10^7$, $10^8$, and $10^9$ samples for $s = 2, 3$, $s = 4, 5$, and $s = 6$, respectively. 
            The physical error probabilities were chosen as $p_{\mathrm{phys}} = 10^{-0.01n - 1.21}$ for integer indices $n = 0, 1, \ldots, 9$, corresponding to logarithmically spaced points between $10^{-1.21}$ and $10^{-1.30}$. 
            To improve visual clarity, a small $s$-dependent horizontal offset was added to each data series.
            The vertical error bars represent 95\% confidence intervals based on the beta distribution.
            The dashed lines represent fitting curves based on the empirical scaling model $p_L = (p_{\mathrm{phys}} / p_{\mathrm{th}})^\alpha$, 
            with fitting parameters $\alpha$ and $p_{\mathrm{th}}$ obtained from the analysis in \cref{fig:logical_error_fitting,fig:pseudo_threshold}.
            Larger size parameters lead to stronger suppression of logical errors.
    }
    \label{fig:logical_error_rates}
\end{figure}

\Cref{fig:logical_error_rates} depicts the result of the numerical evaluation of the logical error rates as a function of the physical error probability. 
This figure demonstrates the suppression of logical errors with an increase in the size parameter $s$.
In particular, combining the BP and order statistics decoding enables us to eliminate the error floor in the low-error regime reported in Ref.~\cite{meister2024efficientsoftoutputdecoderssurface}.

\begin{figure}
    \centering
    \includegraphics[width=1.0\linewidth]{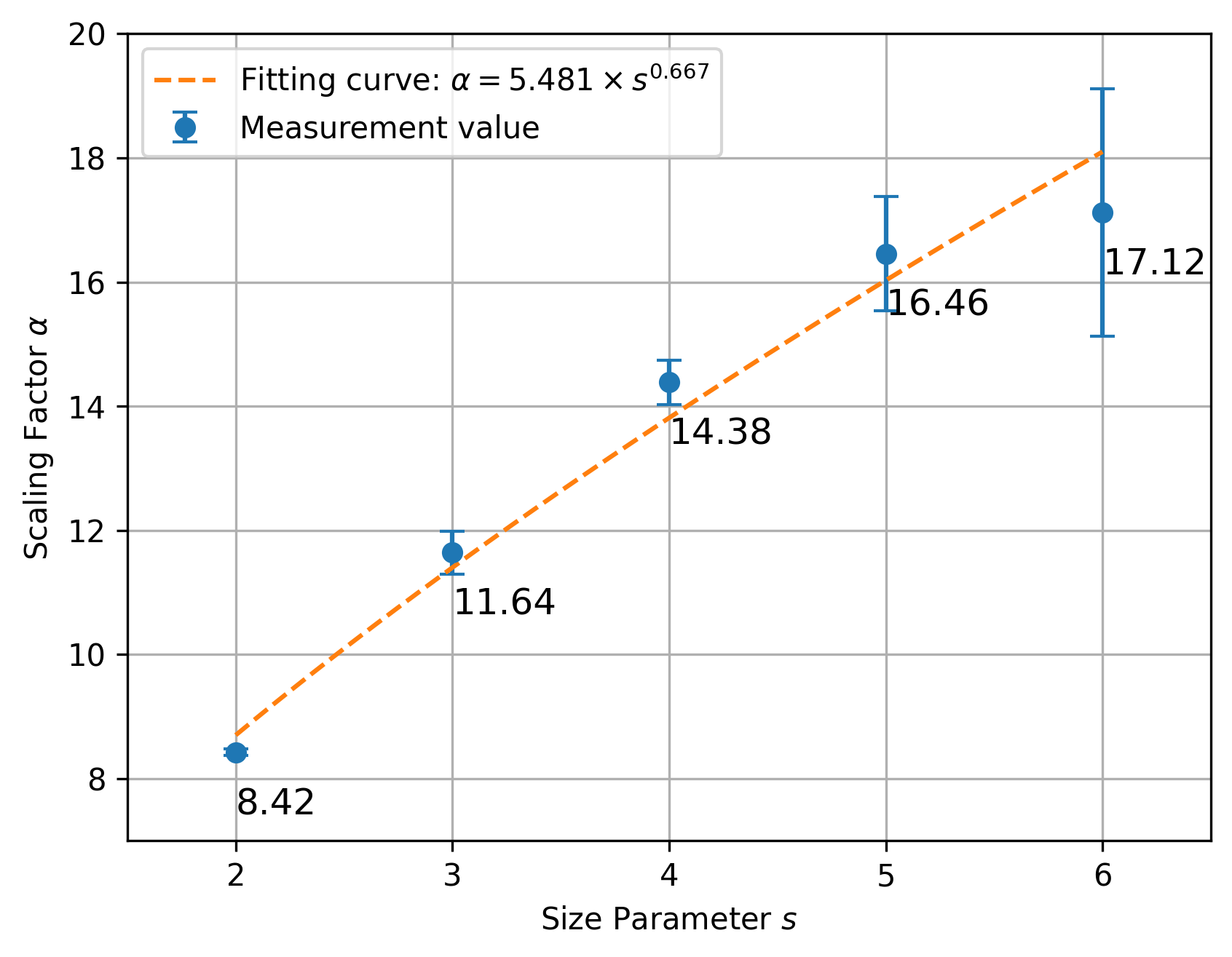}
    \caption{
        Scaling factor $\alpha$ as a function of the size parameter $s$.
        Blue points with error bars show numerical estimates with $1\sigma$ uncertainties obtained from the covariance matrix of the fit, while the orange dashed line represents the power-law fit $\alpha = 5.481\times s^{0.667}$.
        The increase of $\alpha$ with $s$ demonstrates the scalability of the concatenated code.
    }
    \label{fig:logical_error_fitting}
\end{figure}

\begin{figure}
    \centering
    \includegraphics[width=1.0\linewidth]{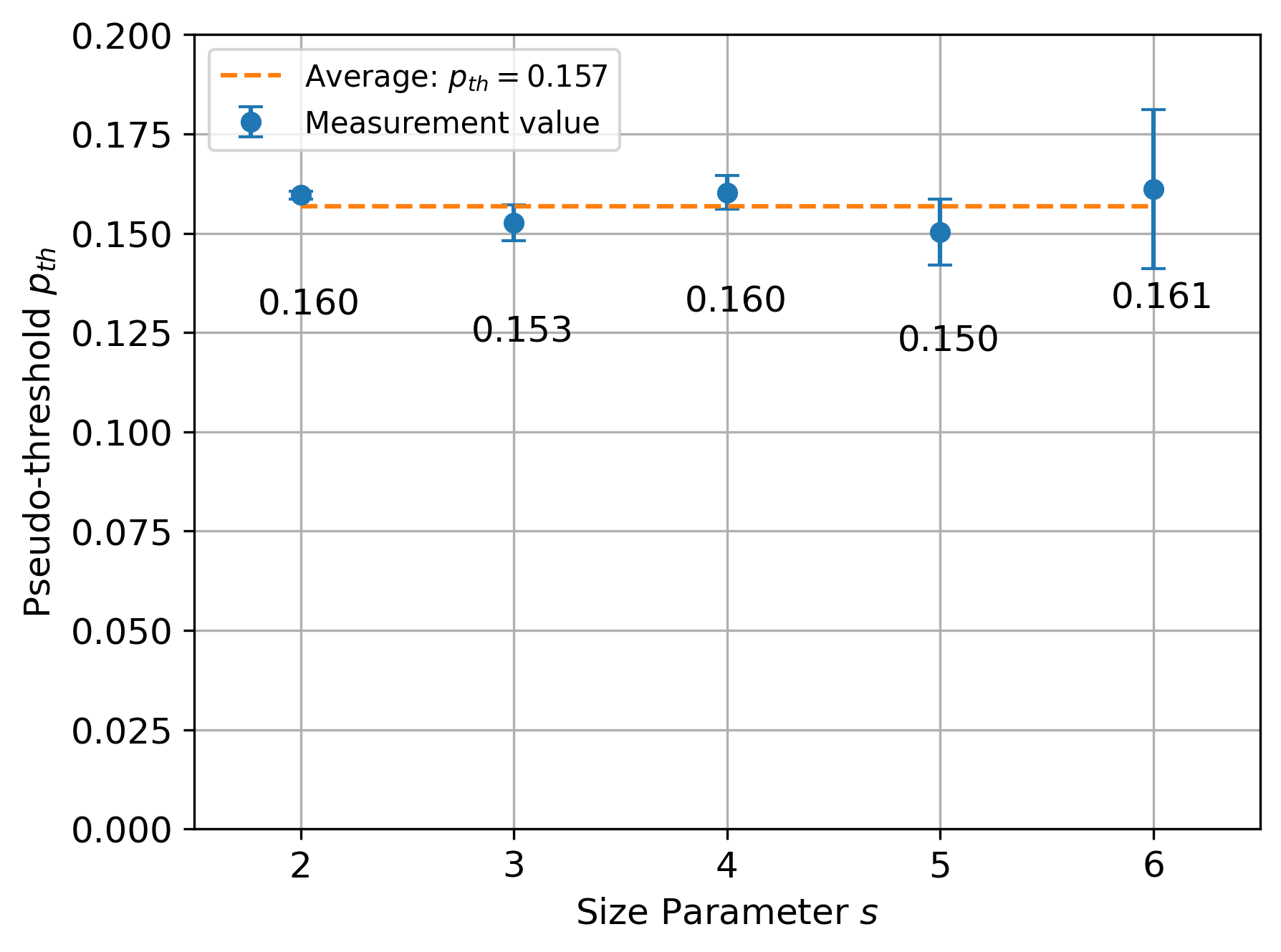}
    \caption{
       Pseudo-threshold $p_\mathrm{th}$ as a function of the size parameter $s$. Blue points with error bars represent numerical estimates with $1\sigma$ uncertainties calculated via the delta method from the covariance matrix of the fit of \cref{eq:logical_error_scaling_formula}, while the orange dashed line indicates the average value $\bar{p}^\mathrm{c}_\mathrm{th} = 0.157$. The results show that the pseudo-threshold remains nearly constant across different $s$.
    }
    \label{fig:pseudo_threshold}
\end{figure}

The scaling of the logical error rate below the pseudo-threshold is modeled as:
\begin{align}
\label{eq:logical_error_scaling_formula}
    p_L = \qty(\frac{p}{p_{\mathrm{th}}})^\alpha,
\end{align}
where $p_{\mathrm{th}}$ and $\alpha$ represent the pseudo-threshold and scaling factor, respectively. 
The fitting values for $\alpha$ and $p_\mathrm{th}$ are shown in \cref{fig:logical_error_fitting,fig:pseudo_threshold}, respectively.
Here, the error bars of $\alpha$ are given by the $1\sigma$ uncertainties obtained from the diagonal entries of the covariance matrix of the fit, 
while those of $p_\mathrm{th}$ are computed using the delta method applied to the same covariance matrix. 
Furthermore, we fit $\alpha$ itself to the formula $\alpha = b\cdot s^c$, obtaining:
\begin{align}
\label{eq:scaling_of_error_scaling_factor}
    b = 5.481 \pm 0.416,\quad c = 0.667\pm 0.055,
\end{align}
where the uncertainties of $b$ and $c$ also correspond to the $1\sigma$ values derived from the covariance matrix of this nonlinear fit.

These results form the basis for the analysis in the next section, where we explore conditions under which the concatenated code outperforms the rotated surface code.

\medskip
\noindent \underline{Comparison with Hard-Decision Decoding}
\medskip

\begin{figure}[ht]
    \centering
    \includegraphics[width=1.0\linewidth]{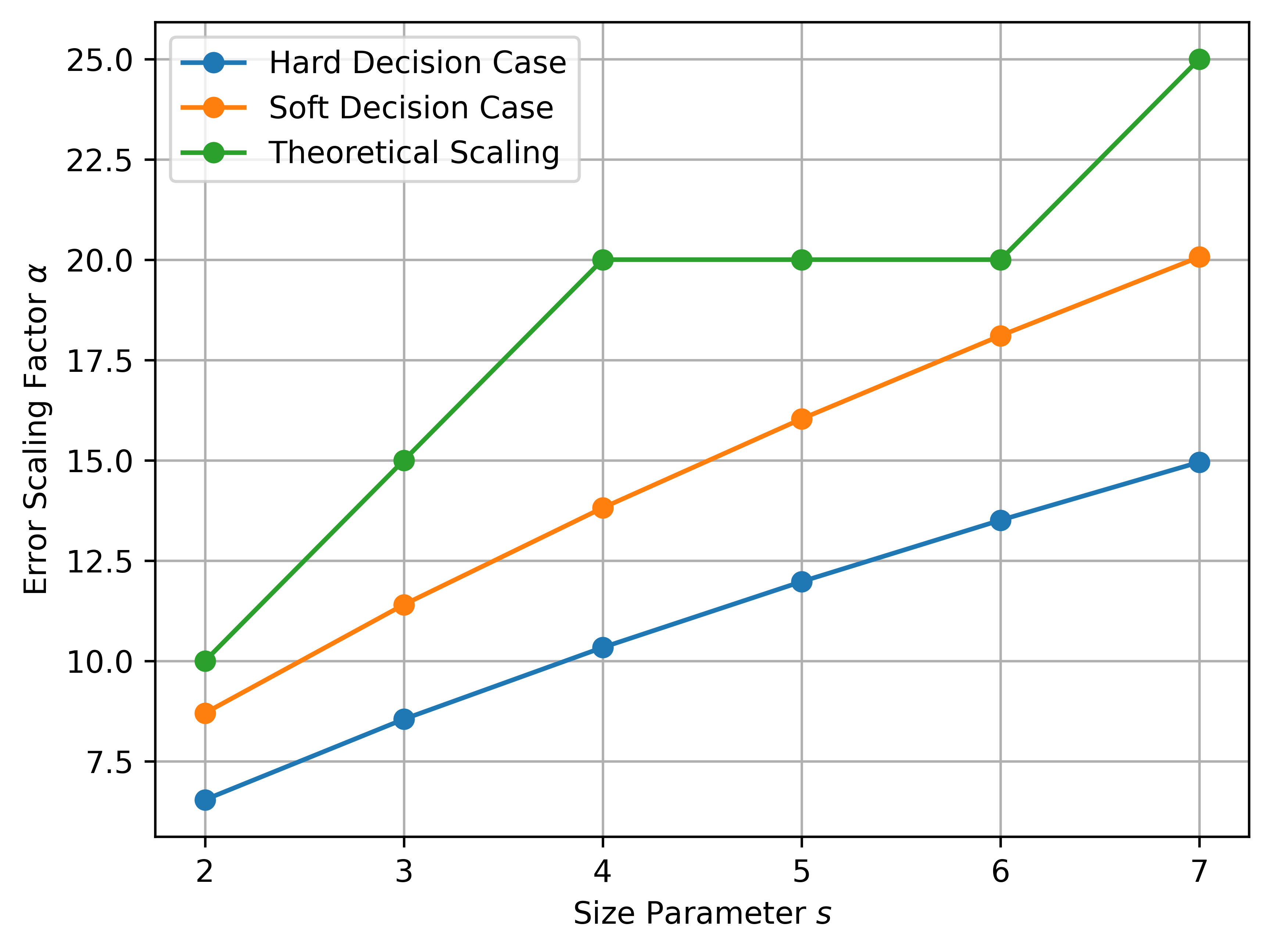}
    \caption{Comparison of error scalings.
    The blue/orange/green lines correspond to the hard-decision/soft-decision/theoretical cases.}
    \label{fig:error_scaling_comparison}
\end{figure}

From \Cref{eq:scaling_of_code_distance}, we can compare how soft-decision decoding improves the error scaling factor.
When hard-decision decoding is applied, the scaling factor of the logical error rate for the concatenated code is reduced to:
\begin{align}
    \alpha_{\mathrm{h}} = \left\lfloor \frac{d_1 + 1}{2} \right\rfloor \left\lfloor \frac{d_2 + 1}{2} \right\rfloor,
\end{align}
where $d_1$ and $d_2$ denote the code distances of the inner and outer codes, respectively.

For our concatenated code, the inner code consists of a rotated surface code with $d_1 = 5$, while the outer HGP code has a code distance $d_2 = d_s$.
This yields:
\begin{align}  
\label{eq:scaling_factor_of_hard_decision}
    \alpha_{\mathrm{h}} = \left\lfloor \frac{5 + 1}{2} \right\rfloor \left\lfloor \frac{d_s + 1}{2} \right\rfloor = 3 \frac{d_s}{2}.
\end{align}
For simplicity, we assume that $d_s$ is even (which is true, at least, for $s \leq 7$).
Substituting \Cref{eq:scaling_of_code_distance} into this, the error scaling factor with arbitrary size parameter is approximately given by
\begin{align}
\label{eq:appoximate_error_scaling_factor_hard_decision}
    \alpha_{\mathrm{h}} = 4.14 s^{0.660}.
\end{align}

In \cref{fig:error_scaling_comparison}, we compare the error scaling factors for the hard-decision case (\ref{eq:appoximate_error_scaling_factor_hard_decision}), the soft-decision case (\ref{eq:scaling_of_error_scaling_factor}), and the theoretical value based on the code distance of the concatenated code, $5d_s/2$.
As shown in the figure, the soft-decision decoding significantly improves the error scaling factor compared to the hard-decision decoding.
However, the observed scaling for the soft-decision decoding still falls short of the theoretical limit.
One possible explanation is that the BP-OS decoder, being an approximate inference algorithm, may occasionally produce incorrect error configurations against syndromes arising from high-weight errors.

\subsection{How Large Must the Hierarchical HGP Code Be to Outperform Rotated Surface Code?}
\label{sec:How_Large_Hierarchical_HGP_Code_Can_Outperform_Rotated_Surface_Code}

In this section, we analyze the conditions under which the proposed concatenated code performs better than the rotated surface code.
Specifically, two key criteria must be satisfied:
(i) the concatenated code must require fewer physical qubits per logical qubit, and (ii) the concatenated code must achieve a lower logical error rate.
We discuss the minimum size parameter to satisfy these conditions with both hard- and soft-decision decoding in the following.

\subsubsection*{Hard-Decision Case}
First, we briefly analyze the conditions under which our concatenated code outperforms the rotated surface code when using hard-decision decoding.

\medskip
\noindent \underline{Condition (i): Fewer Physical Qubits}
\medskip

To ensure better qubit efficiency, the concatenated code should use fewer physical qubits per logical qubit compared to the rotated surface code.
The parameters of these codes are given by:
\begin{align}
    \text{Concatenated code:} &\quad [[625s^2, s^2, 5d_s]], \\
    \text{Rotated surface code:} &\quad [[d^2, 1, d]].
\end{align}
The number of physical qubits per logical qubit for each code is:
\begin{align}
    \frac{\text{Physical qubits}}{\text{Logical qubits}} = 
    \begin{cases}
        625, & \text{(Concatenated code)} \\
        d^2, & \text{(Rotated surface code)}
    \end{cases}
\end{align}

For the concatenated code to be more qubit-efficient, the following inequality must hold:
\begin{align}
    625 \leq d^2.
\end{align}
Taking the square root on both sides, we obtain:
\begin{align}
\label{eq:first_condition}
    d \geq 25.
\end{align}
This result indicates that the minimum code distance of the rotated surface code must be at least 25 for the concatenated code to achieve better qubit efficiency.
In particular, this condition is independent of the size parameter $s$ of the concatenated code, which reflects the constant encoding rate of $1/625$ in this architecture.

\medskip
\noindent \underline{Condition (ii): Lower Logical Error Rate} \medskip

Let us consider a requirement of the logical error rate when the physical error rate is sufficiently below the thresholds.
In this situation, this condition is reduced to one in which the concatenated code has larger logical error scaling than the rotated surface code.
While its error scaling with hard decision was discussed in \Cref{eq:scaling_factor_of_hard_decision}, for the rotated surface code of distance $d$, the scaling factor is given by:
\begin{align}
\label{eq:scaling_factor_of_surface_code}
    \alpha_{\mathrm{r}} = \frac{d+1}{2}.
\end{align}

Comparing these two relations \eqref{eq:scaling_factor_of_hard_decision} and \eqref{eq:scaling_factor_of_surface_code}, the concatenated code achieves a lower logical error rate than the rotated surface code if:
\begin{align}
    \alpha_{\mathrm{h}} \geq \alpha_{\mathrm{r}},
\end{align}
which yields:
\begin{align}
\label{eq:lower_logical_error_rate_condition_of_hard_decision}
    3 d_s - 1 \geq d.
\end{align}

\medskip
\noindent \underline{Summary of Conditions in the Hard-Decision Case}
\medskip

Combining these conditions \eqref{eq:first_condition} and \eqref{eq:lower_logical_error_rate_condition_of_hard_decision}, we derive:
\begin{align}
    3 d_s - 1 \geq 25,
\end{align}
which simplifies to:
\begin{align}
    d_s \geq \frac{26}{3} \approx 8.67.
\end{align}
Thus, for practical implementation, we require $d_s \geq 10$.
From the numerical results of the maximum code distances of the HGP codes (\cref{fig:relation_between_size_and_distance}), this requirement is satisfied if the size parameter obeys:
\begin{align}
    s \geq 7.
\end{align}

\subsubsection*{Soft-Decision Case}
Next, let us discuss how the previous analysis is improved by using soft-decision decoding.

\medskip
\noindent \underline{Condition (i): Fewer Physical Qubits} \medskip

For qubit efficiency, we require the same condition as \Cref{eq:first_condition}, namely:
\begin{align}
    d \geq 25.
\end{align}

\medskip
\noindent \underline{Condition (ii): Lower Logical Error Rate} \medskip

To satisfy the second condition, the logical error rate of the concatenated code must be lower than that of the rotated surface code.
We compare their logical error scaling properties with fixing the physical error rate.

The logical error rate of the concatenated code follows:
\begin{align}
\label{eq:concatenated_scaling}
    p_L \sim \qty(\frac{p}{\bar{p}^\mathrm{c}_{\mathrm{th}}})^{b s^c},
\end{align}
where $\bar{p}^\mathrm{c}_{\mathrm{th}} = 0.157$ is the average pseudo-threshold from \cref{fig:pseudo_threshold}, and the scaling factor $\alpha = b s^c$ depends on the size parameter $s$.
In the following analysis, we assume that the fitting formula holds for an arbitrary size parameter exactly.

For the rotated surface code, the logical error rate follows:
\begin{align}
\label{eq:surface_scaling}
    p_L \sim \qty(\frac{p}{\bar{p}^\mathrm{s}_{\mathrm{th}}})^{\frac{d+1}{2}},
\end{align}
where $\bar{p}^\mathrm{s}_{\mathrm{th}} = 0.1776$ is the pseudo-threshold by the improved BP decoder for surface code~\cite{Criger_2018}.

To determine the minimum size parameter $s$ for which the concatenated code achieves a lower logical error rate, we set:
\begin{align}
    \qty(\frac{p}{\bar{p}^\mathrm{c}_{\mathrm{th}}})^{b s^c} \leq \qty(\frac{p}{\bar{p}^\mathrm{s}_{\mathrm{th}}})^{\frac{d+1}{2}}.
\end{align}
Taking the logarithm on both sides, we obtain:
\begin{align}
    b s^c \log\qty(\frac{p}{\bar{p}^\mathrm{c}_{\mathrm{th}}}) \leq \frac{d+1}{2} \log\qty(\frac{p}{\bar{p}^\mathrm{s}_{\mathrm{th}}}).
\end{align}
Solving for $s$, we obtain the threshold size parameter:
\begin{align}
\label{eq:second_condition}
    s_{\mathrm{rs}}(p, d) \coloneqq \qty(
    \frac{d+1}{2b} \frac{\log(p/\bar{p}^\mathrm{s}_{\mathrm{th}})}{\log(p/\bar{p}^\mathrm{c}_{\mathrm{th}})}
    )^{1/c}.
\end{align}

\begin{figure}[ht]
    \centering
    \includegraphics[width=0.9\linewidth]{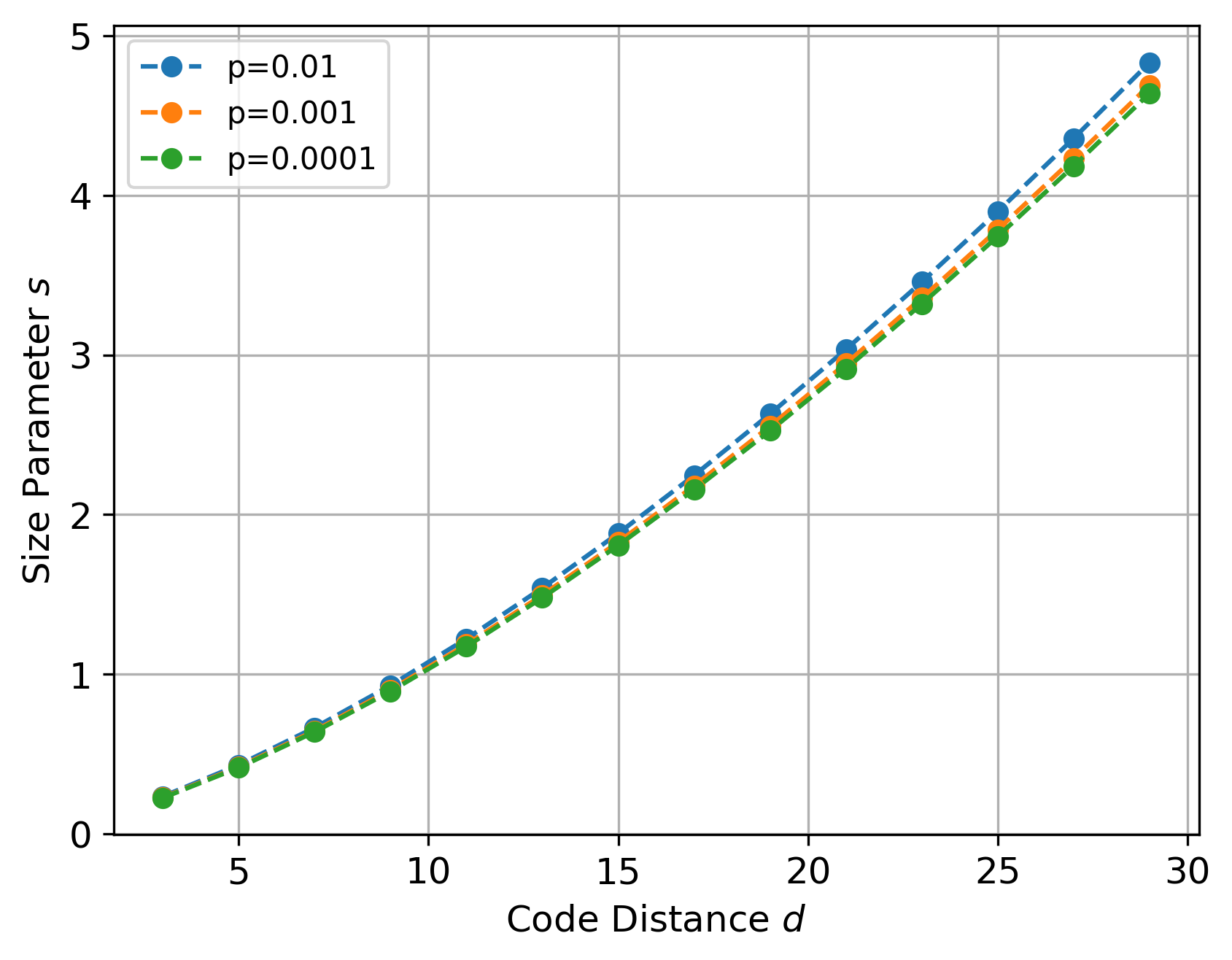}
    \caption{Minimum size parameter $s_{\mathrm{rs}}(p, d)$ required for the concatenated HGP code to achieve a lower logical error rate than the rotated surface code, plotted as a function of the rotated surface code distance $d$ for various physical error probabilities $p = 10^{-2}, 10^{-3},10^{-4}$. The size parameter $s_{\mathrm{rs}}(p, d)$ shows minimal dependence on the physical error rate when $p$ is below the pseudo-thresholds.}
    \label{fig:size_parameter_vs_distance}
\end{figure}

\Cref{fig:size_parameter_vs_distance} illustrates $s_{\mathrm{rs}}(p, d)$ as a function of the rotated surface code distance $d$ for different physical error rates $p=10^{-2},10^{-3},10^{-4}$.

\medskip
\noindent \underline{Summary of Conditions in the Soft-Decision Case}
\medskip

Combining \Cref{eq:first_condition,eq:second_condition}, the concatenated code outperforms the rotated surface code if:
\begin{align}
    s \geq s_{\mathrm{rs}}(p,d) \quad \text{and} \quad d \geq 25.
\end{align}

\begin{figure}[ht]
    \centering
    \includegraphics[width=0.9\linewidth]{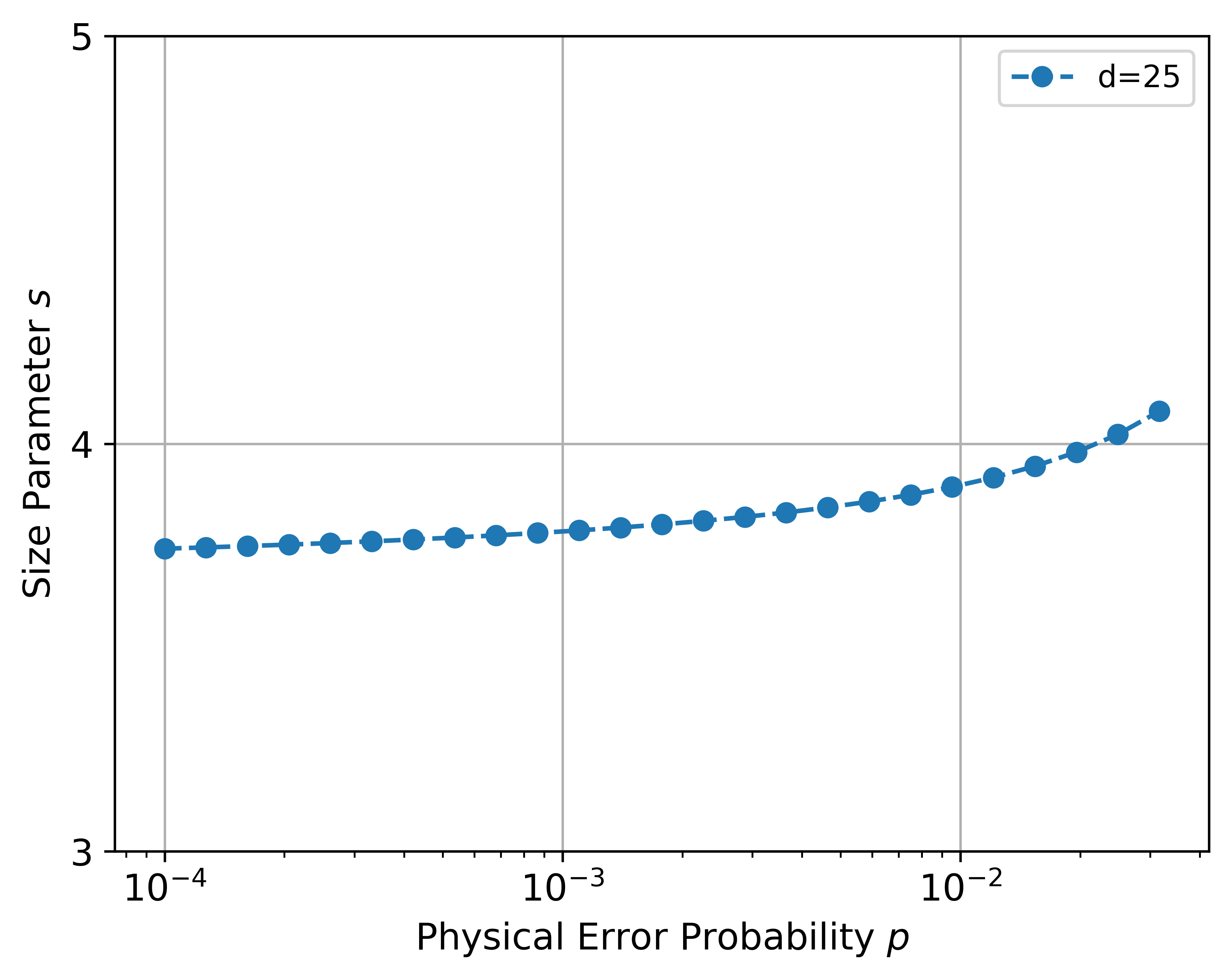}
    \caption{Minimum size parameter $s_{\mathrm{rs}}(p, d = 25)$ required for the concatenated HGP code to achieve a lower logical error rate than the rotated surface code, plotted as a function of the physical error probability $p$.
    If $p$ is around or smaller than $10^{-2}$, $s_{\mathrm{rs}}(p, d = 25)$ stabilizes around $s \simeq 4$, indicating minimal dependence on the physical error rate in this regime.}
    \label{fig:size_parameter_vs_error_rate}
\end{figure}

From \cref{fig:size_parameter_vs_distance}, we observe that $s_{\mathrm{rs}}(p, d)$ shows small dependence on the physical error rate $p$ when $p$ is below the pseudo-thresholds $\bar{p}^\mathrm{c}_{\mathrm{th}}$ and $\bar{p}^\mathrm{s}_{\mathrm{th}}$.
This follows from the small relative difference between the logarithms of these thresholds:
\begin{align}
    \abs{
        \frac{
            \log(\bar{p}^\mathrm{s}_{\mathrm{th}}) - \log(\bar{p}^\mathrm{c}_{\mathrm{th}})
        }{
        \log(p) - \log(\bar{p}^\mathrm{c}_{\mathrm{th}})
        }
    } \ll 1.
\end{align}
Thus, the ratio of logarithms approaches unity:
\begin{align}
    \frac{\log(p/\bar{p}^\mathrm{s}_{\mathrm{th}})}{\log(p/\bar{p}^\mathrm{c}_{\mathrm{th}})} \simeq 1.
\end{align}
As a result, $s(p, d)$ can be approximated by:
\begin{align}
     s(p, d) \simeq \biggl(\frac{d+1}{2b}\biggr)^{1/c}.
    \label{eq:approximated_s}
\end{align}
\Cref{fig:size_parameter_vs_error_rate} shows the dependence of $s_{\mathrm{rs}}(p, d = 25)$ on the physical error rate $p$.
As seen from this figure and \Cref{eq:approximated_s}, $s_{\mathrm{rs}}(p, d = 25)$ becomes approximately constant below the pseudo-thresholds ($\bar{p}^\mathrm{c}_\mathrm{th},\bar{p}^\mathrm{s}_\mathrm{th}$), given by:
\begin{align}
    s(p, d = 25) \simeq \biggl(\frac{13}{b}\biggr)^{1/c} = 3.65\ldots.
\end{align}
Thus, we conclude that our concatenated code can outperform the rotated surface code if
\begin{align}
    s \geq 4 \quad \text{and} \quad d \geq 25.
\end{align}

\begin{figure}[ht]
    \centering
    \includegraphics[width=0.9\linewidth]{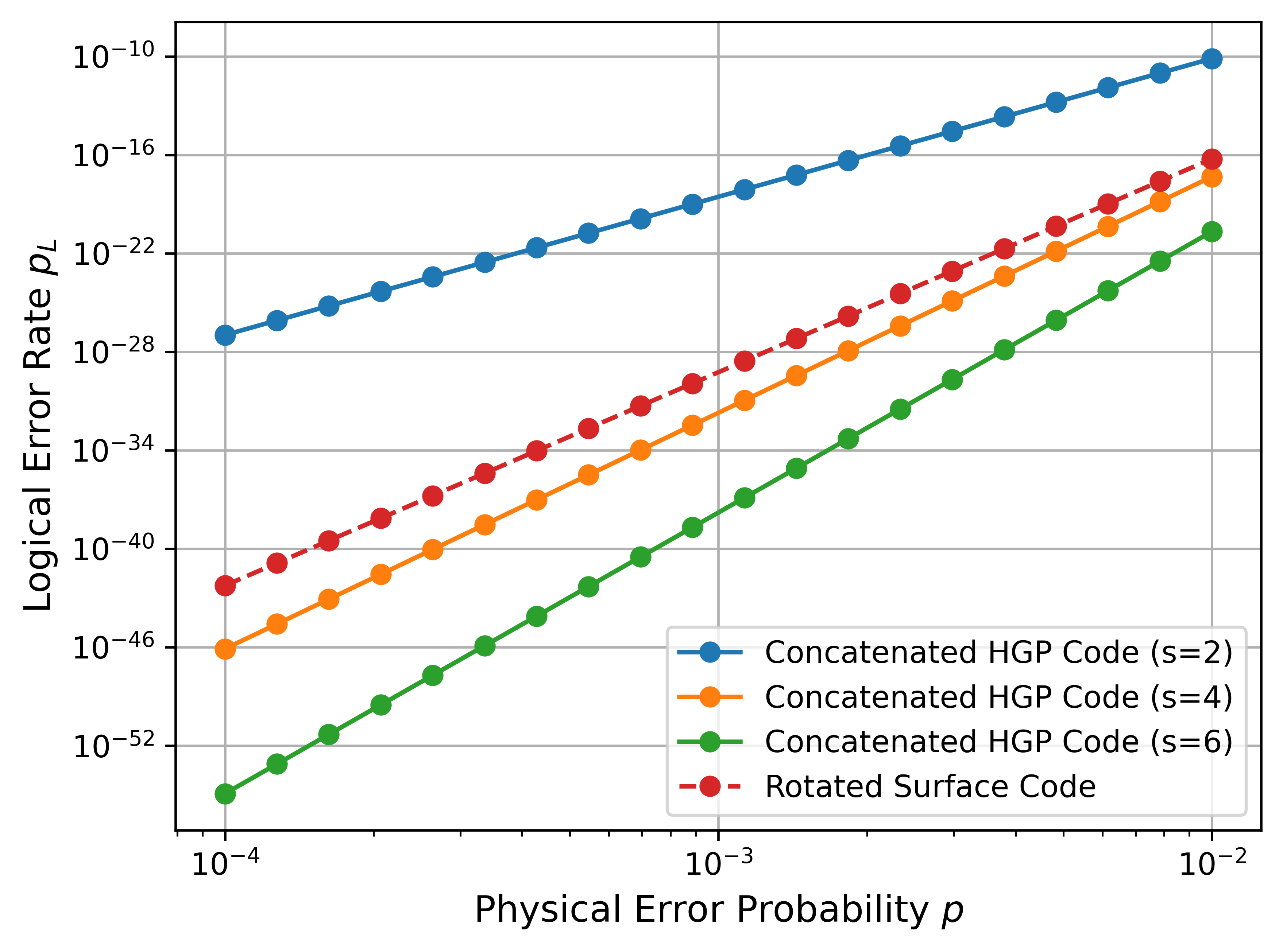}
    \caption{
    Comparison of logical error rates $p_L$ as a function of physical error probability $p$ for concatenated HGP codes with different size parameters $s = 2, 4, 6$ and the rotated surface code with distance $d = 25$, based on the approximate scaling formulae (\ref{eq:concatenated_scaling}), (\ref{eq:surface_scaling}).
    The concatenated codes with $s\geq 4$ exhibit significantly improved logical error suppression compared to the rotated surface code.}
    \label{fig:logical_error_rate_comparison}
\end{figure}

If $s=4$, the concatenated code has the parameters:
\begin{align}
    [[10^4, 16, 2\alpha]], \quad \alpha \simeq 14.38,
\end{align}
whereas the rotated surface code with $d = 25$ has:
\begin{align}
    [[625, 1, 25]].
\end{align}
These results demonstrate that, within practical physical error rates ($p\lesssim 10^{-2}$), the hierarchical code offers a viable alternative to the rotated surface code as a QEC code, providing significantly higher encoding rates while maintaining comparable logical error suppression.
In \cref{fig:logical_error_rate_comparison}, we plot the logical error rate of the concatenated codes with the size parameters $s=2,4,6$ and the rotated surface code of the size $d=25$.

To increase the number of logical qubits, two approaches can be considered. 
One option is to prepare multiple independent copies of the concatenated code, while the other is to increase the size parameter $s$ of the HGP code. 
From an encoding rate perspective, both methods are equivalent, as they preserve the constant encoding rate of the hierarchical code.
However, in terms of logical error suppression, increasing the size parameter is more advantageous, since it directly enhances the code distance, as indicated by \Cref{eq:scaling_of_error_scaling_factor}.

Beyond the specific case of HGP codes, the choice of outer qLDPC codes significantly impacts the performance of the hierarchical scheme.
The threshold code distance required for the concatenated code to outperform is determined by the square root of the inverse of the encoding rate, $\sqrt{(1/625)^{-1}}$. Using a qLDPC code with a higher encoding rate lowers this threshold, allowing the concatenated code to surpass the rotated surface code with fewer physical qubits.
There is also another option of changing the inner code, and we discuss how the situation changes if we change the size of the rotated surface code of the inner code in \Cref{app:generalization_to_any_rotated_suface_code}.

In addition, logical error suppression depends on the scaling properties of the qLDPC code. A qLDPC code with a lower logical error rate reduces the minimum required size parameter $s$ for outperforming the rotated surface code. The HGP code, despite its structured construction, is not an asymptotically good qLDPC code, as its code distance scales only as the square root of the number of physical qubits, even when constructed from optimized classical codes. Incorporating qLDPC codes with better distance scaling, such as those where the code distance scales linearly or sublinearly with the number of physical qubits, could further enhance the hierarchical QEC framework.

\section{Conclusion and Discussion}
\label{sec:summary_outlook}

In this work, we proposed and analyzed a hierarchical quantum error correciton (QEC) approach based on the concatenation of hypergraph product (HGP) codes with the rotated surface code.
This construction provides a practical method for implementing quantum low-density parity-check (qLDPC) codes on planar quantum architectures, such as superconducting qubits, where nearest-neighbor connectivity is typically required.

In this paper, we demonstrate that concatenated HGP codes maintain a constant encoding rate while achieving sublinear growth in code distance with respect to the size parameter $s$.
This contrasts with surface codes, whose encoding rate decreases as the code size increases.
Next, numerical simulations under a depolarizing noise model showed that the hierarchical code achieves a lower logical error rate than the rotated surface code when the physical error rate $p$ is around or smaller than $10^{-2}$ ($p\lesssim 10^{-2}$).
Finally, we derived conditions under which the concatenated code outperforms the rotated surface code as a QEC code in terms of both qubit efficiency and logical error rate.
Specifically, for a size parameter $s \geq 4$ (which yields the number of logical qubits $k\geq 16$) and code distance $d \geq 25$, the hierarchical code exhibits superior performance with parameters $[[10^4, 16, 2\alpha]]$, where $\alpha \simeq 14.38$.
These results imply the potential of concatenated HGP codes as a scalable and resource-efficient solution for fault-tolerant quantum computing (FTQC).
The ability to achieve high encoding rates while maintaining strong error correction capabilities is particularly valuable in early fault-tolerant quantum computers, where physical qubits are still a limited resource.

While these results imply the potential of our hierarchical code as a quantum error correction scheme, we emphasize that our study does not yet constitute a complete proposal for FTQC.
In particular, the simulations were performed under a code capacity noise model, which neglects important factors such as measurement errors and circuit-level noise.
As such, our aim was not to evaluate the full practicality of the scheme for FTQC, but rather to identify a regime in which the proposed hierarchical codes can outperform surface codes as a QEC code in terms of logical error suppression and qubit efficiency.
To establish the viability of this approach for FTQC, further studies are required to incorporate overheads from syndrome extraction circuits, ancilla qubits, and time delays due to lattice surgery.
Nonetheless, we believe that our results provide a valuable foundation for exploring hierarchical QEC architectures as a potentially more resource-efficient alternative to conventional surface-code-based schemes.

Several challenges remain to be addressed for practical implementation.
In this study, we employed a decoding strategy that combines a lookup-table decoder for the rotated surface code with a BP-OS decoder for the HGP code.
While this hybrid approach enabled accurate decoding in our setting, further optimization—particularly in the context of soft-decision decoding—is crucial for enhancing performance at larger code sizes. Developing more efficient and scalable decoding algorithms thus remains an important open problem.

In particular, we chose a rotated surface code of size $L = 5$ as the inner code to allow exact computation of the syndrome-conditioned logical error probabilities via a precomputed lookup table.
If we increase the code distance beyond $L = 5$, it rapidly becomes impractical due to the exponential growth of the syndrome space and the resulting memory requirements.
To overcome this limitation, recent advances in soft-output decoding—such as the complementary gap method~\cite{Hutter_2014,gidney2023yokedsurfacecodes,Bomb_n_2024,meister2024efficientsoftoutputdecoderssurface}—provide efficient approximations of these probabilities.
Incorporating such techniques into hierarchical decoding schemes offers a promising avenue for supporting larger inner codes while maintaining high decoding accuracy.

Moreover, our analysis was limited to a code capacity noise model, which assumes independent depolarizing noise and neglects measurement errors, as we mentioned above.
Extending the study to more realistic noise environments—such as those with correlated or biased noise, or with faulty measurements—is essential for assessing the practical performance of hierarchical codes.
Finally, integrating these codes into complete fault-tolerant quantum computing architectures, including protocols for logical gate implementation and state preparation, is a critical direction for future research.

In conclusion, the concatenation of HGP codes with rotated surface codes represents a promising strategy to advance fault-tolerant quantum computation. 
By addressing these challenges, hierarchical QEC architectures could play a key role in the development of large-scale quantum systems with reduced resource requirements and enhanced error resilience.

\begin{acknowledgments}
This work is supported by the MEXT Quantum Leap Flagship Program (MEXT Q-LEAP) Grant No.~JPMXS0120319794, JST COI-NEXT Grant No.~JPMJPF2014, and JST CREST JPMJCR24I3.
\end{acknowledgments}

\section*{Data Availability}
The data that support the findings of this article are openly available~\cite{haruna_2025_15660988}~(\href{https://doi.org/10.5281/zenodo.15660987}{DOI 10.5281/zenodo.15660987}).

\appendix

\section{Generalization to Arbitrary inner Surface Code Distance in Hard-Decision Decoding}
\label[appendix]{app:generalization_to_any_rotated_suface_code}

In this appendix, we discuss the condition for our concatenated code to outperform the rotated surface code when we generalize the size of the inner code in hard-decision decoding.

If we change the size of a inner rotated surface code to an arbitrary odd distance $L_1$, the error scaling factor of the concatenated code with the hard-decision decoding is changed to:
\begin{align}
    \alpha_{\mathrm{h}} = \frac{L_1+1}{2} \frac{d_s}{2}.
\end{align}
Repeating the arguments in \Cref{sec:How_Large_Hierarchical_HGP_Code_Can_Outperform_Rotated_Surface_Code}, \Cref{eq:lower_logical_error_rate_condition_of_hard_decision} generalizes to:
\begin{align}
\label{eq:lower_logical_error_rate_condition_of_hard_decision_in_general_size}
    \frac{L_1+1}{2} d_s -1 \geq d.
\end{align}

The condition on the number of physical qubits also changes since the inner surface code now has parameters $[[L_1^2,1,L_1]]$.
The requirement for fewer physical qubits is then:
\begin{align}
    L_1^2 \cdot 25 \leq d^2.
\end{align}
Solving for $d$, we obtain:
\begin{align}
\label{eq:fewer_physical_qubit_condition_of_hard_decision_in_general_size}
    d \geq 5L_1.
\end{align}
Combining \Cref{eq:lower_logical_error_rate_condition_of_hard_decision_in_general_size,eq:fewer_physical_qubit_condition_of_hard_decision_in_general_size}, we obtain:
\begin{align}
    \frac{L_1+1}{2} d_s -1 \geq 5L_1.
\end{align}
Solving for $d_s$, we get:
\begin{align}
    d_s \geq 2 \frac{5L_1+1}{L_1+1}.
\end{align}
When $L_1$ is equal to $3$, the right-hand side of this equality is given by 8.
For $L_1 \geq 5$, it is monotonically increasing and satisfies:
\begin{align}
    8 < 2\frac{5L_1+1}{L_1+1} < 10.
\end{align}
Thus, the condition for the hard-decision decoding case simplifies to:
\begin{align}
    d_s \geq 
    \begin{cases}
        8  \quad & (L_1=3) \\
        10 \quad & (L_1\geq 5)
    \end{cases}
\end{align}
Again, we assume that $L_1$ is an odd integer and $d_s$ is even.
This requirement yields that the size parameter obeys:
\begin{align}
    s \geq 
    \begin{cases}
        4  \quad & (L_1=3) \\
        7 \quad & (L_1\geq 5)
    \end{cases}
\end{align}

\bibliography{ref}
\bibliographystyle{unsrt}

\end{document}